\documentclass[journal]{IEEEtran}
\usepackage{amssymb}
\usepackage{amsmath,amsfonts}
\usepackage{algorithmic}
\usepackage{array}
\usepackage{textcomp}
\usepackage{stfloats}
\usepackage{url}
\usepackage{verbatim}
\usepackage{graphicx}
\usepackage{graphicx}
\usepackage{ascmac}

\usepackage{bm}
\usepackage{amsmath}
\usepackage{amssymb}
\usepackage{amsfonts}
\usepackage{color}
\usepackage{wrapfig}
\usepackage{url}
\usepackage{siunitx}
\usepackage{mathtools}

\usepackage{diagbox}
\usepackage{enumerate}
\usepackage{comment}
\usepackage{balance}

\ifCLASSINFOpdf

\else

\fi

\begin{document}

\title{Stability analysis for large-scale multi-agent molecular communication systems}

\author{Taishi~Kotsuka,~
        Yutaka~Hori,~\IEEEmembership{Member,~IEEE}

    \thanks{This work was supported in part by JSPS KAKENHI Grant Numbers JP21H05889, JP22J10554, and JP23H00506, and in part by JST SPRING Grant Number JPMJSP2123.}

    \thanks{T. Kotsuka and Y. Hori are with the Department of Applied Physics and Physico-Informatics, Keio University, Kanagawa 223-8522 Japan. Correspondence should be addressed to Y. Hori (email: tkotsuka@keio.jp; yhori@appi.keio.ac.jp).}

}

\maketitle

\begin{abstract}
Molecular communication (MC) is recently featured as a novel communication tool to connect individual biological nanorobots. It is expected that a large number of nanorobots can form large multi-agent MC systems through MC to accomplish complex and large-scale tasks that cannot be achieved by a single nanorobot. However, most previous models for MC systems assume a unidirectional diffusion communication channel and cannot capture the feedback between each nanorobot, which is important for multi-agent MC systems. In this paper, we introduce a system theoretic model for large-scale multi-agent MC systems using transfer functions, and then propose a method to analyze the stability for multi-agent MC systems. The proposed method decomposes the multi-agent MC system into multiple single-input and single-output (SISO) systems, which facilitates the application of simple analysis technique for SISO systems to the large-scale multi-agent MC system. Finally, we demonstrate the proposed method by analyzing the stability of a specific large-scale multi-agent MC system and clarify a parameter region to synchronize the states of nanorobots, which is important to make cooperative behaviors at a population level.
\end{abstract}

\begin{IEEEkeywords}
Molecular communications, Feedback control, Transfer function, Biological system modeling, Diffusion.
\end{IEEEkeywords}

\IEEEpeerreviewmaketitle

\section{Introduction}
\label{sec:intro}

In nature, bacteria are known to communicate with each other using diffusion-based molecular communication (MC) to achieve cooperative functions such as biofilm formation. Recently, many researchers attempted to use this MC as a novel communication tool to connect individual biological nanorobots (artificial cells) in vivo \cite{tatsuya2018molecular,Bi2021,Soldner2020,lotter2023,Farsad2016}. In particular, a large number of dispersed nanorobots can form a large multi-agent system through MC, which is expected to enable nanorobots to accomplish complicated and large-scale tasks, leading to engineering applications such as targeted drug delivery \cite{Femminella2015,Gao2014}. In order to design such a multi-agent MC system that can achieve the desired performance, it is important to 
construct a dynamical model for a large-scale multi-agent MC system and analyze its fundamental properties.

\smallskip
\par
An important feature that needs to be incorporated into the model of multi-agent MC systems is the information sharing and the feedback of signals between nanorobots. 
Dynamic models of one-to-one and one-to-$n$ MC systems were constructed based on diffusion equations, and many communication properties were analyzed \cite{Pierobon2010,Chude-Okonkwo2015,Huang2021,Lotter2020,maximilian2020,Akdeniz2017}. In \cite{Pierobon2010}, a system theoretic model of a MC system which consists of a sender nanorobot and a receiver nanorobot was constructed and the frequency response characteristics of the MC channel were analyzed based on Green's function of diffusion equation. However, since these models assume unidirectional MC channels, they cannot be easily extended to models for $n$-to-$n$ multi-agent systems, which require bidirectional MC channels for feedback. A model capturing the feedback between each nanorobot was proposed by the authors' group \cite{Hara2021}, and its stability was analyzed. However, the previously proposed stability analysis method \cite{Hara2021} was applicable only to the MC systems with two nanorobots.  
On the other hand, there are several studies with large-scale multi-agent MC systems considering the reaction in the nanorobots based on the reaction-diffusion (RD) equation \cite{Kotsuka2022,yutaka2015coordinated,Hsia2012,Kashima2015}. However, RD-based MC models assume that the length between the nanorobots is close enough so that the disruption of the signal by the MC channel can be ignored, and thus, the model is not suitable for the analysis and design of the multi-agent MC systems composed of a population of distributed nanorobots for applications in vivo such as drug delivery. This motivates us to develop a more versatile model and analysis methods for large-scale multi-agent MC systems for broader engineering applications of MC systems.

\par
\smallskip
In control engineering, many analysis and design methods were developed for multi-agent mechanical robots by capturing them as feedback systems of agents and communication channels. Since high-speed communication such as electrical or optical communication are generally considered in such multi-agent systems, the communication channels are modeled as a constant gain matrix by ignoring the dynamics of the communication \cite{Qin2017,Cao2013,Olfati-Saber2007}. However, in multi-agent MC systems, the delay in the communication channels is not an ignorable factor because the diffusion of signal molecules is not fast enough compared to the reaction kinetics in nanorobots (agents). In particular, MC is known to significantly attenuate signals in the high-frequency band due to the smoothing effect of diffusion \cite{Pierobon2010,kotsuka2023}. This implies that the unique frequency-dependent delay characteristic of the diffusion communication channel should be explicitly considered for better analysis and design of MC systems. 
From a control engineering perspective, it is therefore a new challenge to construct system theoretic models and methods to control large-scale multi-agent systems with communication channels that have diffusion-based dynamics.

\smallskip
\par
In this paper, we first propose a system theoretic model for large-scale multi-agent MC systems that consist of a feedback of local reaction systems inside nanorobots and communication channels modeled by a diffusion equation. We then develop a method to analyze the stability of large-scale multi-agent MC systems, one of the fundamental properties for the analysis and design of control systems, using transfer functions. 
Specifically, we first approximately model large-scale multi-agent MC systems as circulant MC systems that consist of $n$ homogeneous nanorobots with periodic boundaries at the left/right ends of the system. We then obtain a system theoretic model of the circulant MC system, which contains a large-scale transfer function matrix based on the transfer function of bidirectional MC channels previously derived by the authors' group \cite{kotsuka2023}. The proposed method decomposes the transfer function matrix of the circulant MC system into multiple single-input and single-output (SISO) systems in the same spirit as the stability analysis for linear systems with generalized frequency variables \cite{hara2009LTI}. This decomposition allows for the stability analysis of the large-scale multi-agent MC system by combining simple analysis techniques for SISO systems. Moreover, since the decomposed system captures the information for spatial frequencies of multi-agent MC systems, the proposed method can analyze the response of multi-agent MC systems based on the spatial frequency, which potentially leads to the analysis for the condition for biological pattern formation such as Turing pattern formation \cite{Turing1952}. Finally, we demonstrate the proposed method by analyzing the stability of a specific large-scale multi-agent MC system.

\smallskip
\par
This paper is organized as follows. In the next section, we model large-scale multi-agent MC systems as a circulant MC system with periodic boundaries based on a diffusion equation. We then introduce system theoretic models of the circulant MC system expressed as multi-input and multi-output (MIMO) dynamical systems using a transfer function in Section \ref{sec:system}. In Section \ref{sec:stability}, we provide a method to analyze the stability of the circulant MC system by decomposing the system into multiple SISO systems and show the relation between the decomposed systems and spatial frequency. The proposed method is demonstrated for the analysis of a specific large-scale multi-agent MC system in Section \ref{sec:numerical}. Finally, the paper is concluded in Section \ref{sec:conclusion}.

\smallskip
\par
\noindent
{\bf Notations:}
The following notations are used throughout this paper: the set of real values is defined by $\mathbb{R}$, and the set of complex values is defined by $\mathbb{C}$. The superscript is used to represent the dimension of the vector space, {\it e.g.,} $\mathbb{R}^n$. The identity matrix is defined by $I_n\in\mathbb{R}^{n\times n}$. The Laplace transform of a function $z(t)$ is defined by $Z(s) \coloneqq \mathcal{L}\{z(t)\} = \int^{\infty}_{0}z(t)e^{-st}dt$, where $s=\sigma+j\omega$ is the complex variable with the real part $\sigma$ and the imaginary part $\omega$.

\section{Model of multi-agent MC systems and problem formulation}
\label{sec:model}

\subsection{Mathematical models}

\begin{figure}
    \centering
    \includegraphics[width=0.99\linewidth]{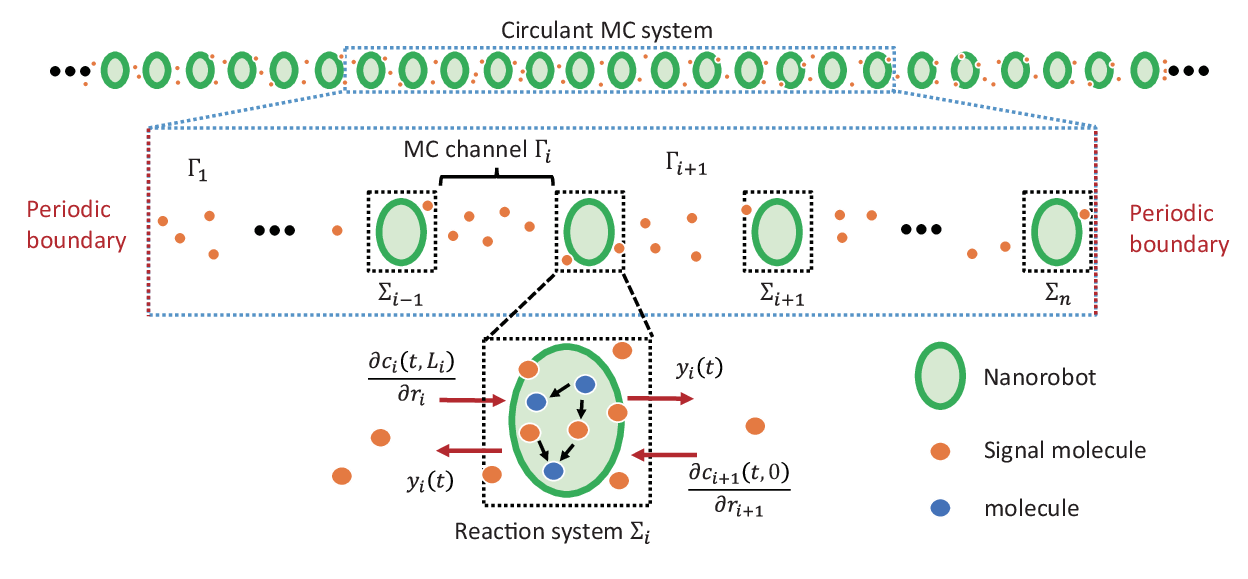}
    \caption{Illustration of multi-agent MC system, where multiple nanorobots communicate with each other via MC channels $\Gamma_i$. The reaction system $\Sigma_i$ captures the reaction of molecules in the nanorobot and the emission/absorption of the signal molecule. }
    \label{fig:mcsystems}
\end{figure}

We consider multi-agent MC systems with countless homogeneous nanorobots in one-dimensional spatial domain, where the nanorobots could be bacterial cells or vesicles that have reaction systems consisting of biomolecules such as genes and proteins as shown in Fig. \ref{fig:mcsystems}. In MC systems, each nanorobot interferes with the state of its neighboring nanorobots by communication through diffusion of signal molecules in the fluidic environment, which in turn affects the state of the entire nanorobot population to achieve cooperative behavior. 
Examples of such systems include a synthetic quorum sensing system, where engineered bacteria synchronize their states via diffusion of acyl-homoserine lactones (AHLs) \cite{danino2010synchronized}, and a distributed biocomputing circuit, where bacterial populations perform as logic gates by using 2,4-diacetylphloroglucinol (DAPG) and salicylate (Sal) as signal molecules \cite{Du2020}. 
In these examples, the signal molecules produced by biomolecular reaction in each nanorobot are emitted from the nanorobot and diffuse in the fluidic environment. %
The signal molecules reaching other nanorobots are absorbed at the surface and become reactants to drive reactions within the nanorobots. 

\par
\smallskip
To model the dynamics of the multi-agent MC system, we consider a large number of nanorobots, say $n$ nanorobots, in one-dimensional space. The left and the right end of the space is connected as shown in Fig. \ref{fig:mcsystems} (blue dotted box), {\it i.e.,} the boundary is periodic, to approximately model the dynamics of countless nanorobots using a small sub-unit in the space (see Fig. \ref{fig:mcsystems}). 
We call this system {\it a circulant MC system}. 
Let $\Sigma_i\,(i=1,2,\cdots,n)$ denote the reaction system that captures the reaction of molecules in the $i$-th nanorobot and the emission/absorption of the signal molecules %
and $\Gamma_i$ denote the $i$-th MC channel between the reaction systems $\Sigma_{i-1}$ and $\Sigma_i$. %
In what follows, the subscript $i$ is taken by modulo $n$, {\it i.e.,} $\Sigma_0 := \Sigma_n$ since periodic boundaries are applied. %

\par
\smallskip
We denote the concentration of the signal molecules at position $r\in[0,L]$ in the MC channel $\Gamma_i$ by $c_i(t,r)$, where $L$ is the communication length of each channel $\Gamma_i$. In practice, the communication length between nanorobots may not be uniform, but we here consider an average length $L$  to approximately model the collective behavior of many nanorobots. 
Since the behavior of the signal molecules in the MC channel follows Fick's law, the dynamics of the concentration $c_i(t,r)$ of the signal molecules can be modeled by the diffusion equation 
\begin{equation}
    \frac{\partial c_i(t,r)}{\partial t} = \mu\frac{\partial^2c_i(t,r)}{\partial r^2},
    \label{eq:diffusion}
\end{equation}
where $\mu$ is the diffusion coefficient. The boundary conditions at $r=0$ and $r=L$ are dynamic Dirichlet boundary conditions defined by 
\begin{align}
c_i(t,0) &= y_{i-1}(t),\\
    c_i(t,L) &= y_{i}(t),
\end{align}
where $y_i(t)$ is the emitted concentration of the signal molecule from the $i$-th reaction system $\Sigma_i$, which is thus determined by the dynamics of the reaction system $\Sigma_i$. %

\par
\smallskip
For many practical examples \cite{yutaka2015coordinated,Collins2000}, the dynamics of the $i$-th reaction system $\Sigma_i$ associated with $m-1$ species of molecules can be modeled by nonlinear state-space models as
\begin{align}
    \displaystyle\frac{d\bm{x}_i(t)}{dt} &= \bm{f}(\bm{x}_i) + \bm{B} w_{i}(t),\\
    y_{i}(t) &= \bm{C}\bm{x}_i(t),\label{eq:Rsystem}
\end{align}
where the state $\bm{x}_i(t)=[x_{i,1}, x_{i,2},\cdots,x_{i,m}]^T$ is the concentrations of the molecules associated with reactions occurring inside and outside of the $i$-th nanorobot, and $x_{i,l}$ is the concentration of the $l$-the molecular species. The variable $x_{i,m}$ represents the concentration of the signal molecule outside of the $i$-th nanorobot. The vector function $\bm{f}(\cdot)$ represents the dynamics of the reactions in the nanorobot and the membrane transport, and $\bm{B}=[0,\cdots,0,1]^T\in\mathbb{R}^{m}$ and $\bm{C}=[0,\cdots,0,1]\in\mathbb{R}^{1\times m}$ are the input and the output vectors, respectively.
The variable $w_i(t)$ is the input fluxes from the MC channels $\Gamma_i$ and $\Gamma_{i+1}$ to the $i$-th reaction system $\Sigma_i$ following the Fick's first law and the conservation law for the total number of the signal molecule. Specifically, $w_i(t)$ is
\begin{equation}
w_{i}(t) = \frac{\mu}{\Delta r}\left(\frac{\partial c_{i+1}(t,0)}{\partial r} - \frac{\partial c_i(t,L)}{\partial r}\right),\label{eq:flux}
\end{equation}
where $\Delta r$ is the size of the reaction system $\Sigma_i$. The output $y_i(t)$ is the concentration of the signal molecule outside of the nanorobot. 

\smallskip
\par
\noindent
{\bf Remark 1.}
The balance of the time scale between the reaction and diffusion kinetics is important for MC systems. 
A typical molecular communication of biological systems occurs between nanorobots that are apart about $1-1000\,\si{\micro m}$ \cite{Burmeister2020}. 
Within this length, the difference of the time scale between intracellular genetic circuit and the extracellular diffusion volumes is not very large. The reaction rate of biomolecular systems such as gene regulatory networks can be effectively captured by the degradation rate of molecules, which is typically between $10^{-2}\,\mathrm{min^{-1}}$ and $10^0\,\mathrm{min^{-1}}$ \cite{Basu2005}. On the other hand, the characteristic rate of diffusion of signal molecules is calculated by $\mu/L^2$. Assuming that the communication length $L$ is around $10^2\,\si{\micro m}$, the characteristic time of diffusion becomes the same order of magnitude with the reaction rate since the diffusion coefficient $\mu$ of signal molecules is typically between $10^0\,\si{\micro m^2\cdot min^{-1}}$ and $10^3\,\si{\micro m^2\cdot min^{-1}}$ \cite{Bi2021}. Thus, the dynamics of MC systems are affected by both of the reaction and the diffusion kinetics, which motivates us to analyze MC systems based on models capturing both of the reaction and the diffusion.

\subsection{Problem formulation}

A potential application of the multi-agent MC system is to synchronize the states of nanorobots, which is important to make cooperative behaviors at a population level. 
In fact, the circulant MC system can converge to a spatially homogeneous equilibrium point at steady states by designing the reaction systems and the MC channels appropriately. In other words, the population of nanorobots can synchronize their state against small perturbation around an equilibrium point at steady state. %
To analyze the convergence to the synchronized states, an important concept is {\it the stability} of the system. Specifically, the states of nanorobots converge to the homogeneous equilibrium point if and only if the MC system is asymptotically stable around the homogeneous equilibrium point. In what follows, we introduce the definitions of the stability to develop mathematically rigorous stability analysis methods in the following sections.

\medskip
\par
\noindent
{\bf Definition 1.}
Consider the system (\ref{eq:diffusion}) -- (\ref{eq:flux}). A constant vector $\bm{x}^*\in\mathbb{R}^m$ is a spatially homogeneous equilibrium point if 
\begin{enumerate}
    \item $\bm{f}(\bm{x}^*)=0$, and  %
    \item $c_i(t,r)=x^*_m$ for all $i=1,2,\cdots,n$ and for all $r\in[0,L]$, where $x^*_m$ is the $m$-th entry of $\bm{x}^*$.%
\end{enumerate}

\medskip
\par
\noindent
{\bf Assumption 1.}
The circulant MC system (\ref{eq:diffusion}) -- (\ref{eq:flux}) has a spatially homogeneous equilibrium point $\bm{x}^*$.

\medskip
\par
If the biomolecular reaction system within nanorobots has an equilibrium point, the circulant MC system satisfies Assumption 1 since (2) of Definition 1 is automatically satisfied when $\bm{f}(\bm{x}^*)=0$. 
Next, we show the definition of the stability and the asymptotic stability of the circulant MC system (\ref{eq:diffusion}) -- (\ref{eq:flux}) around a spatially homogeneous equilibrium point $\bm{x}^*$. %

\medskip
\par
\noindent
{\bf Definition 2.}
Consider the system (\ref{eq:diffusion}) -- (\ref{eq:flux}), and %
let $\bm{x}^*$ %
be a spatially homogeneous equilibrium point. The spatially homogeneous equilibrium point is locally stable if the following 1 and 2 hold:
\begin{enumerate}
    \item For all $\epsilon_1>0$, if there exists $\delta_1>0$ such that $||\bm{x}_i(0)-\bm{x}^*||<\delta_1$, then $||\bm{x}_i(t)-\bm{x}^*||<\epsilon_1$ for all $t\geq0$ and $i=1,2,\cdots,n$. %
    \item For all $\epsilon_2>0$, if there exists $\delta_2>0$ such that $\int^L_0{||c_i(0,r)-x_m^*||dr}<\delta_2$, then $\int^L_0{||c_i(t,r)-x_m^*||dr}<\epsilon_2$ for $t\geq0$ and all $i=1,2,\cdots,n$.
\end{enumerate}

\smallskip
\par
\noindent
{\bf Definition 3.}
Consider the system (\ref{eq:diffusion}) -- (\ref{eq:flux}), and %
let $\bm{x}^*$ %
be a spatially homogeneous equilibrium point. The spatially homogeneous equilibrium point is locally asymptotically stable if it is locally stable, and the following 1 and 2 hold:
\begin{enumerate}
    \item If there exists $\delta_1>0$ such that $||\bm{x}_i(0)-\bm{x}^*||<\delta_1$, then $\lim_{t\rightarrow\infty}{||\bm{x}_i(t)-\bm{x}^*||}=0$ for $i=1,2,\cdots,n$.%
    \item If there exists $\delta_2>0$ such that $\int^L_0{||c_i(0,r)-x_m^*||dr}<\delta_2$, then $\lim_{t\rightarrow\infty}{\int^L_0{||c_i(t,r)-x_m^*||dr}}=0$ for $i=1,2,\cdots,n$.
\end{enumerate}

\par
\smallskip
In the next section, we first express the circulant MC system as a multi-input multi-output (MIMO) dynamical system using the transfer functions derived from the diffusion equation (\ref{eq:diffusion}) and the state-space model (\ref{eq:Rsystem}). We then show that the MIMO circulant MC system can be decomposed into multiple single-input single-output (SISO) systems to facilitate the stability analysis.

\section{System theoretic model for circulant MC systems}
\label{sec:system}

\begin{figure}
    \centering
    \includegraphics[width=0.90\linewidth]{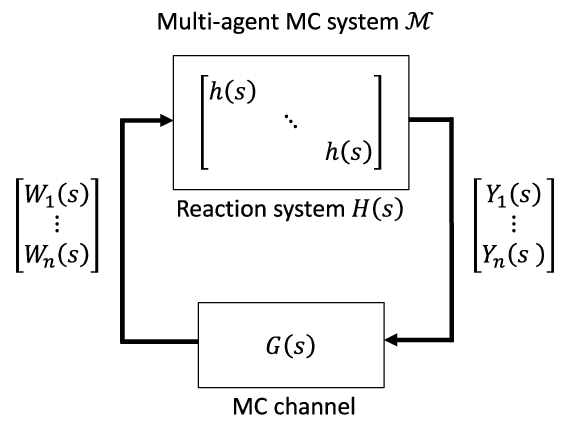}
    \caption{The block diagram of the multi-agent MC system $\mathcal{M}$.}
    \label{fig:mimo}
\end{figure}

In this section, we first derive the transfer function of the circulant MC system since the stability of systems can be analyzed by examining the roots of the denominator of the transfer functions. We then show the MIMO representation of the MC system based on the transfer functions. 

\par
\smallskip
Transfer functions represent the ratio of the Laplace transform of the input/output signals and allow us to analyze and design the system in the complex domain, where many tools in control theory can be applied. 
Let $Y_i(s)\coloneqq\mathcal{L}[y_i(t)]$ and $W_i(s)\coloneqq \mathcal{L}[w_i(t)]$ denote the Laplace transform of the signals $y_i(t)$ and $w_i(t)$, respectievly. Specifically, $W_i(s)$ is written as
\begin{equation}
    W_i(s) = \frac{\mu}{\Delta r}\left(\frac{\partial C_{i+1}(s,0)}{\partial r} - \frac{\partial C_i(s,L)}{\partial r} \right),
    \label{eq:ws}
\end{equation}
where $C_i(s,\cdot)\coloneqq\mathcal{L}[c_i(t,\cdot)]$. 
Then, the transfer function of the reaction system $\Sigma_i$ is defined by
\begin{equation}
    h(s) \coloneqq \frac{Y_i(s)}{W_i(s)} = \bm{C}(sI_m-A)^{-1}\bm{B},\label{eq:Treaction}
\end{equation}
where $A$ is the Jacobian matrix of $\bm{f(\bm{x})}$ at the homogeneous equilibrium point $\bm{x}^*$ \cite{astrom2008}. 
The transfer function $h(s)$ represents the relation between the input $w_i(t)$ and the output $y_i(t)$ of the reaction system $\Sigma_i$ in the complex domain. 

\par
\smallskip
We next consider the transfer function of the MC channel $\Gamma_i$. The input-output relation between the concentrations $y_{i-1}(t)=c_i(t,0)$ and $y_i(t)=c_i(t,L)$, and the concentration gradients $\partial c_i(t,L)/\partial r$ and $\partial c_i(t,L)/\partial r$ can be expressed using the Laplace transform of these variables as
\begin{align}
    &\frac{\partial C_i(s,0)}{\partial r} = \frac{1}{2}g_1(s) C_i(s,0) + g_2(s) C_i(s,L),\label{eq:Tc0}\\
    &\frac{\partial C_i(s,L)}{\partial r} = -g_2(s) C_i(s,0) - \frac{1}{2}g_1(s) C_i(s,L),\label{eq:TcL}
\end{align}
where 
\begin{align}
    g_1(s) &\coloneqq -\sqrt{\frac{s}{\mu}} \frac{2}{\tanh{\left(\frac{L}{\sqrt{\mu}}\sqrt{s}\right)}},\label{eq:Gtan}\\ %
    g_2(s) &\coloneqq \sqrt{\frac{s}{\mu}}\frac{1}{\sinh{\left(\frac{L}{\sqrt{\mu}}\sqrt{s}\right)}}. %
    \label{eq:Gsin}
    \end{align}
Equations (\ref{eq:Tc0}) -- (\ref{eq:TcL}) are derived by the Laplace transform of the diffusion equation (\ref{eq:diffusion}) with the Dirichlet boundary conditions at $r=0$ and $r=L$ \cite{kotsuka2023}.

\par
\smallskip
Based on the input-output relations (\ref{eq:Treaction}) -- (\ref{eq:TcL}), the circulant MC system consisting of the MC channels $\Gamma_i$ and the reaction systems $\Sigma_i$ can be represented as the MIMO system $\mathcal{M}$:
\begin{equation}
\begin{split}
    \bm{Y}(s)
    &= H(s) \bm{W}(s),\\
    \bm{W}(s)
    &= G(s) \bm{Y}(s),
\end{split}\label{eq:mimo}
\end{equation}
where $\bm{Y}(s):=[Y_1(s),Y_2(s),\cdots,Y_n(s)]^T$ and $\bm{W}(s):=[W_1(s),W_2(s),\cdots,W_n(s)]^T$. The transfer function matrix $H(s)$ is the diagonal matrix as
\begin{equation}
    H(s) = h(s)I_n \label{eq:H}
\end{equation}
representing the reaction systems. The transfer function matrix $G(s)$ is the circulant tridiagonal matrix representing the MC channels defined by 
\begin{equation}
    G(s) \coloneqq \frac{\mu}{\Delta r}\left[\begin{matrix}
        g_1(s)&g_2(s)&0&\cdots&0&g_2(s)\\g_2(s)&g_1(s)&g_2(s)&\cdots&0&0\\0&g_2(s)&\ddots&\ddots&0&\vdots \\\vdots&\ddots&\ddots&\ddots&\ddots&0 \\0&\ddots&\ddots&\ddots&\ddots&g_2(s) \\g_2(s)&0&\cdots&0&g_2(s)&g_1(s)
    \end{matrix} \right].\label{eq:Gsp}
\end{equation}
The function $g_1(s)$ represents the transmission characteristics of the signal molecules from the emitting nanorobot to itself. In other words, $g_1(s)$ characterizes the process where the emitted signal molecules undergo extracellular diffusion and eventually come back to the emitting nanorobot itself. The function $g_2(s)$ represents the transmission characteristics of the signal from the emitting nanorobot to the neighbor, and it has a low-pass characteristics whose cut-off frequency decreases with the communication length $L$ increases, which is consistent with intuition about diffusion phenomena \cite{kotsuka2023}. 
The block diagram of the MIMO circulant MC system is shown in Fig. \ref{fig:mimo}. %

\smallskip
\par
\noindent
{\bf Remark 2.}
Equations (\ref{eq:Gtan}) and (\ref{eq:Gsin}) show the dependency of the frequency characteristics of the MC system on the communication length $L$. Suppose the length $L$ of an MC system is large, and the characteristic rate of diffusion is sufficiently small so that the intracellular reactions occur on a much shorter time scale. In the limit of infinite length $L\rightarrow\infty$, the transfer functions $g_1(s)$ and $g_2(s)$ are asymptotic to
\begin{align}
    \lim_{L\rightarrow\infty}g_1(s) &= -\sqrt{\frac{s}{\mu}},\\
    \lim_{L\rightarrow\infty}g_2(s) &= 0,
\end{align}
and thus the transfer function matrix $G(s)$ becomes a diagonal matrix. Therefore, the MC system is decomposed into the independent systems consisting of the reaction system $h(s)$ and $g_1(s)$, which represents the self-feedback kinetics due to the take up of molecules that are emitted from the nanorobot itself. This implies that the stability of the MC system mainly depends on the stability of the independent systems when the rate of intracellular reactions is large enough.

\par
\smallskip
The MIMO circulant MC system differs from conventional multi-agent mechanical robots in that the communication channel is modeled by the transfer function matrix $G(s)$, which represents the diffusion dynamics, rather than by a constant matrix. 
Based on the system theoretic model for the MIMO circulant MC system (\ref{eq:mimo}) with the transfer functions, one can analyze the stability, robustness, and performance of the system in the complex domain. However, these analyses become difficult when the MIMO circulant MC system $\mathcal{M}$ consists of a large number of nanorobots, {\it i.e.,} $n$ is large since it requires computing the determinant of a large non-rational matrix. In the next section, we propose the method to analyze the stability of the MIMO circulant MC system $\mathcal{M}$ by reducing the problem of computing the determinant of a large matrix to that of computing multiple non-rational scalar equations.

\section{Stability analysis for circulant MC systems}
\label{sec:stability}

In this section, we show a method to analyze the stability of the circulant MC system. Specifically, we first show that the stability analysis of the MIMO circulant MC system with $n$ nanorobots can be reduced to that of $n$ SISO systems by some linear transformation. We then discuss the physical interpretation of the decomposed SISO system based on spatial frequency. 

\subsection{Stability analysis by decomposition of the MC system}
\label{sec:stabilityDFT}

We first show a necessary and sufficient stability condition for the closed-loop system (\ref{eq:mimo}) based on the characteristic equation. The characteristic equation $p(s)=0$ of the closed-loop system (\ref{eq:mimo}) is defined by
\begin{equation}
    p(s) := \det(I_n-G(s)H(s)) = 0 \label{eq:character}
\end{equation}
(Definition 4.6 of reference \cite{skogestad2005multivariable}). It is known in control engineering that the closed-loop (\ref{eq:diffusion}) -- (\ref{eq:flux}) is locally asymptotically stable if and only if all roots of the characteristic equation $p(s)=0$ lie in the open left half plane (OLHP) of the complex plane (Theorem 5.7 of reference \cite{robust1996}).

\par
\smallskip
The roots of the characteristic equation $p(s)=0$ are not easy to analyze since $p(s)$ is defined by the determinant of a matrix with non-rational functions. 
In what follows, we show a theorem that the stability analysis for the MIMO system (\ref{eq:mimo}) can be reduced to that for $n$ SISO systems using the property of the circulant matrix $G(s)$. %

\par
\medskip
\noindent
{\bf Theorem 1.}
Consider the circulant MC system (\ref{eq:diffusion}) -- (\ref{eq:flux}) and suppose Assumption 1 holds. Let $\bm{x}^*$ denote a spatially homogeneous equilibrium point of the circulant MC system. The following three statements are equivalent.
\begin{enumerate}
    \item The spatially homogeneous equilibrium point $\bm{x}^*$ is locally asymptotically stable.
    \item All roots of $p(s)=0$ lie in the open left half plane of the complex plane, {\it i.e.,} $\mathrm{Re}[s]<0$.
    \item All roots of the characteristic equations
    \begin{equation}
    \hat{p}_i(s) := 1-\lambda_i(s)h(s) = 0\label{eq:chat}
\end{equation}
lie in the OLHP of the complex plane for all $i$, where 
\begin{align}
i = 
\begin{cases}
    1,2,\cdots,\frac{n}{2}+1 & \mathrm{when\ }n\mathrm{\ is\ even}\\
    1,2,\cdots,\frac{n+1}{2} & \mathrm{when\ }n\mathrm{\ is\ odd}
\end{cases}
\end{align}
and
\begin{equation}
    \lambda_i(s) = g_1(s) + 2g_2(s)\cos{ \left(2\pi\frac{i-1}{n}\right)}. %
    \label{eq:lambda}
\end{equation} 
\end{enumerate}

\par
\medskip
\noindent
{\bf Proof.}
The equivalence of the statement 1 and 2 is well-known in control engineering (Theorem 5.7 of reference \cite{robust1996}), and thus we prove the equivalence of the statements 2 and 3. 
Since the MC channel $G(s)$ is the circulant matrix, $G(s)$ can be diagonalized by the discrete Fourier transform (DFT) matrix $F\in\mathbb{C}^{n\times n}$ \cite{thetransform2001} as 
\begin{equation}
    \Lambda(s) := FG(s)F^*,
\end{equation}
where $\Lambda(s)$ is the diagonal matrix $\si{diag}(\lambda_1(s), \lambda_2(s), \cdots, \lambda_n(s))$ and
\begin{equation}
    F = \frac{1}{\sqrt{n}}\left[\begin{matrix}
        1&1&1&\cdots&1\\1&\alpha&\alpha^2&\cdots&\alpha^{n-1} \\1&\alpha^{2}&\alpha^4&\cdots&\alpha^{2(n-1)} \\\vdots&\vdots&\vdots&\ddots&\vdots \\1&\alpha^{(n-1)}&\alpha^{2(n-1)}&\cdots&\alpha^{(n-1)(n-1)}
    \end{matrix} \right]\label{eq:DCT}
\end{equation}
with $\alpha = e^{-j2\pi/n}$. 
Using the DFT matrix $F$, the characteristic equation (\ref{eq:character}) can be transformed as
\begin{align}
    p(s) &= \det\left(F^*(I_n-FG(s)F^*FH(s)F^*)F\right)\nonumber\\
    &= \det(I_n-h(s)\Lambda(s))\nonumber\\
    &= \prod_i^n[1-\lambda_i(s)h(s)] = \prod_i^n\hat{p}_i(s)=0,\label{eq:charatrans}
\end{align}
where we use $FH(s)F^*=h(s)I_n$ and $\det(F)\det(F^*)=1$ because $F$ is the unitary matrix. The theorem holds since Eq. (\ref{eq:charatrans}) shows that the roots of the characteristic equation $p(s)=0$ coincide with those of $\hat{p}_i(s)=0$ for all $i$. %

\par
\begin{figure}
    \centering
    \includegraphics[width=0.5\linewidth]{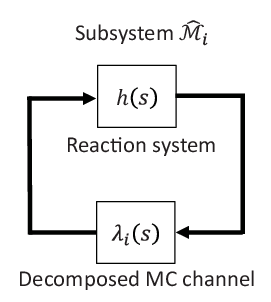}
    \caption{Decomposed multi-agent MC system $\hat{\mathcal{M}}_i$.}
    \label{fig:blocksub}
\end{figure}

\medskip
Theorem 1 shows that the problem of solving the determinant of the non-rational matrix (\ref{eq:character}) can be reduced to that of solving non-rational scalar equation (\ref{eq:charatrans}). Theoretically, this means that MIMO system $\mathcal{M}$ with the $n\times n$ circulant matrix $G(s)$ can be decomposed into $n$ SISO systems $\hat{\mathcal{M}}_i$ as shown in Fig. \ref{fig:blocksub}, which facilitates the analysis for the stability of the MC system using methods developed in control engineering. %

\subsection{Physical interpretation of decomposed systems}
\label{sec:sfreq}

Next, we discuss the physical interpretation of the decomposed system $\hat{\mathcal{M}}_i$ shown in Fig. \ref{fig:blocksub} based on spatial frequency. 
We consider the decomposed subsystem $\hat{\mathcal{M}}_i$ shown in Fig. \ref{fig:blocksub}, whose input-output relation is given by
\begin{equation}
\begin{array}{cc}
    &\hat{Y}_i(s) = h(s)\hat{W}_i(s),\\
    &\hat{W}_i(s) = \lambda_i(s)\hat{Y}_i(s),
    \end{array}\label{eq:decomposedsystem}
\end{equation}
where
\begin{equation}
\begin{array}{cc}
    &\displaystyle \hat{Y}_i(s) = \bm{F}_iY(s) = \frac{1}{\sqrt{n}}\sum_{l=1}^{n}Y_l(s)\alpha^{(l-1)(i-1)}, \\
    &\displaystyle \hat{W}_i(s) = \bm{F}_i^TW(s) = \frac{1}{\sqrt{n}}\sum_{l=1}^{n}W_l(s)\alpha^{(l-1)(i-1)}, \label{eq:fourierU}
    \end{array}
\end{equation}
and $\bm{F}_i\in\mathbb{C}^{1\times n}$ is the $i$-th row of the DFT matrix $F$. Equation (\ref{eq:fourierU}) corresponds to the discrete Fourier transform of the signal $\bm{Y}(s)$ and $\bm{W}(s)$, and thus $\hat{Y}_i(s)$ and $\hat{W}_i(s)$ represent the Fourier component with a spatial frequency $\xi_i=2\pi(i-1)/nL$ of the signals $\bm{Y}(s)$ and $\bm{W}(s)$, respectively. 
This means that the decomposed subsystem $\hat{\mathcal{M}}_i$ consisting of $h(s)$ and $\lambda_i(s)$ expresses the dynamics of the Fourier component of the signal with a specific spatial frequency $\xi_i=2\pi(i-1)/nL$. Therefore, the stability of the subsystem $\hat{\mathcal{M}}_i$ implies that the MC system converges to the spatially homogeneous equilibrium state when a perturbation with a component of spatial frequency $\xi_i$ is input to the system. Thus, the proposed method is able to analyze for which Fourier component of perturbations the circulant MC system converges to the spatially homogeneous equilibrium point by analyzing the stability of the decomposed subsystems.

\par
\smallskip
Next, we consider the circulant MC system with a large number of $n$ by considering the limit of $n\rightarrow\infty$. 
For the infinite number of nanorobots $n\rightarrow\infty$, Eq. (\ref{eq:fourierU}) implies that the dynamics of the subsystems $\hat{\mathcal{M}}_i$ is asymptotic to the sum of the Fourier components for all continuous spatial frequencies $\xi$. 
We here show that the stability analysis for decomposed subsystems $\hat{\mathcal{M}}_i$ when $n\rightarrow\infty$ boils down to the problem of computing a non-rational scalar equation for a certain range of the parameter. 

\par
\medskip
\noindent
{\bf Lemma 1.}
Consider the characteristic equation (\ref{eq:chat}) and define
\begin{equation}
\mathbb{X}_n\coloneqq\left\{s\in\mathbb{C} \mid  \hat{p}_i(s)=0,\, i=1,2,\cdots,n \right\}
\end{equation}
and
\begin{equation}
\bar{\mathbb{X}}\coloneqq\left\{s\in\mathbb{C} \mid \hat{p}(s) = 0,\, \lambda(s) \coloneqq g_1(s) + \beta g_2(s),\, \beta\in[-2,2] \right\},
\end{equation}
where $\hat{p}(s) := 1-\lambda(s)h(s)$. 
The following two statements hold:
\begin{enumerate}
    \item $\mathbb{X}_n\subset\bar{\mathbb{X}}$.%
    \item In the limit of $n\rightarrow\infty$, $\mathbb{X}_n$ is asymptotic to $\bar{\mathbb{X}}$.
\end{enumerate}

\par
\medskip
Lemma 1 is derived since the cosine function in the second term of Eq. (\ref{eq:lambda}) varies continuously for $i$ in the range from $-1$ to $1$ by taking the limit $n\rightarrow\infty$. Lemma 1 shows that the characteristic equations (\ref{eq:chat}) for $i=1,2,\cdots,n$ is asymptotic to the characteristic equation $\hat{p}(s)=0$ for all $\beta$ by $n\rightarrow\infty$, 
and thus it implies that the stability of the circulant MC system for $n\rightarrow\infty$ can be analyzed by computing the roots of the characteristic equation $\hat{p}(s)=0$ for all $\beta$. %

\smallskip
\par
\noindent
{\bf Remark 3.}
A potential application of Theorem 1 is the analysis of biological pattern formation such as Turing pattern formation. 
Turing patterns are spatially periodic patterns of molecular concentrations formed by cell populations communicating via signal molecules. It is expected that if Turing patterns can be artificially formed/designed by MC systems, the function of each nanorobot can be changed according to its positional information. The mechanism of Turing pattern formation can be elucidated by the spatial frequency-based analysis for a reaction-diffusion (RD) equation \cite{Turing1952}. 
An RD system can be decomposed into multiple subsystems, which express the dynamics of Fourier components for spatial frequencies $\xi$. It is known that 
if the subsystem with the spatial frequency $\xi=0$ is stable and one or more of the subsystems with nonzero spatial frequencies $\xi_u$ is unstable, a spatial pattern with a period corresponding to the spatial frequency $\xi_u$ is formed when spatial white noise is input. This is because only the dynamics of the unstable Fourier component with $\xi_u$ do not converge to an equilibrium point, and thus the molecular concentration appears periodically in space corresponding to the spatial frequency $\xi_u$. 
In the same way, considering the circulant MC system, if the subsystem $\hat{M}_1(s)$ is stable and one or more of the subsystems $\hat{M}_{i}(s)\,(i=2,3,\cdots,n)$ is unstable, the spatially periodical pattern would emerge. By developing the proposed method, the parameter conditions for the Turing pattern formation with the circulant MC system might be derived. 

\subsection{Graphical stability test}
\label{sec:graph}

In the previous section, we have shown the method to reduce the problem of computing the determinant of a large matrix to that of computing multiple non-rational scalar equations for the stability analysis of the MIMO circulant MC system $\mathcal{M}$. However, finding all the roots of a non-rational scalar equation containing exponential functions is not easy since there are an infinite number of roots in some cases. We here show one of the useful methods to graphically analyze if all the roots of the characteristic equation (\ref{eq:chat}) lie in the OLHP of the complex plane. To this end, we first show the definition of the Nyquist contour, which is used for the graphical stability test.

\par
\smallskip
\noindent
{\bf Definition 4.} A contour is the Nyquist contour if it consists of the imaginary axis from $0-j\infty$ to $0+j\infty$ and a semicircular arc with a radius $R\rightarrow \infty$ starting at $0+j R$ and traveling clockwise to $0-jR$.

\par
\smallskip
\noindent
{\bf Lemma 2.} Consider the characteristic equation (\ref{eq:chat}) %
and let $P$ denote the number of the roots of the equation $\mathrm{den}(h(s))=0$ whose real part is positive, where $\mathrm{den}(\cdot)$ is the denominator of the function. All roots of the characteristic equation (\ref{eq:chat}) lie in the OLHP of the complex plane 
if and only if the Nyquist plot of the open-loop transfer function $-\lambda_i(j\omega)h(j\omega)$ makes $P$ anti-clockwise encirclements of the point -1 and does not pass through the point -1 in the complex plane, where the Nyquist plot is the trajectory of a function along the Nyquist contour. %

\par
\smallskip

Lemma 2 is based on the Nyquist stability criterion \cite{skogestad2005multivariable}. The proof is in \ref{sec:lemma2proof}. In summary, the procedure of the proposed method to analyze the stability of the circulant MC system is as follows. First, we apply Theorem 1 to decompose the stability analysis of the circulant MC system $\mathcal{M}$ into that of $n$ subsystems $\hat{\mathcal{M}}_i$. We then use Lemma 2 to analyze the stability of the decomposed subsystem $\hat{\mathcal{M}}_i$.

\par
\medskip
\noindent
{\bf Remark 4.}
Theorem 1 and Lemma 2 provide the necessary and sufficient condition for the local stability of the multi-agent MC system $\mathcal{M}$, which has the channel $G(s)$ with fractional-order transfer functions, rather than a constant matrix. The proposed method captures the MC system using the similar approach for the stability analysis for linear systems with generalized frequency variables and the communication channel modeled by a constant gain matrix 
\cite{hara2009LTI}, and decomposes the transfer function matrix $G(s)$ of the circulant MC system into multiple SISO systems in the same spirit as the decomposition method of the circulant constant gain matrix proposed in \cite{Hori2011}. Extending these previous results, the proposed method enables to analyze the stability of the multi-agent MC system $\mathcal{M}$. 
%
%

\begin{comment}
\begin{equation}
    \xi_i = \frac{i}{nL}
\end{equation}

\par
\smallskip
[The number of nanorobots $n$ and $\alpha$]
\begin{eqnarray}
    \alpha = e^{-j2\pi\frac{i-1}{n}} + e^{-j2\pi\frac{(i-1)(n-1)}{n}}
\end{eqnarray}
\begin{equation}
    -2\leq\alpha\leq2
\end{equation}
for $i=1,2,\cdots,n$ and $n\geq3 \in \mathcal{Z}$.
Therefore, by computing Eq. (12) for $-2\leq\alpha\leq2$ we can analyze the stability of circulant structured MC systems in any number of nanorobots.
\end{comment}

\section{Numerical example}
\label{sec:numerical}

In this section, we demonstrate the stability analysis for a specific example of the circulant MC system. In particular, we use Theorem 1 and Lemma 2 to analyze the stability of the circulant MC system $\mathcal{M}$ 
for showing that the stability of system $\mathcal{M}$ depends on the communication length $L$ and the diffusion coefficient $\mu$. We then show that the state of the circulant MC system either converges to the spatially homogeneous equilibrium state or transitions to another equilibrium state depending on the spatial frequency of small perturbations added to the system. 

\subsection{Stability analysis for the circulant MC system with an activator-repressor-diffuser genetic circuit}

\begin{figure}
    \centering
    \includegraphics[width=0.99\linewidth]{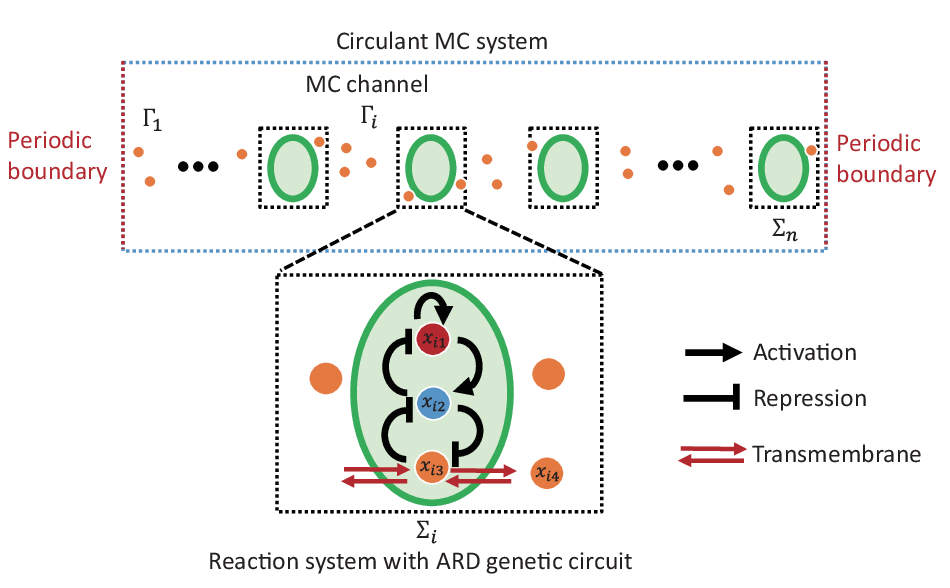}
    \caption{The circulant MC system with ARD genetic circuit. }
    \label{fig:ard}
\end{figure}

\begin{figure}
    \centering
    \includegraphics[width=0.99\linewidth]{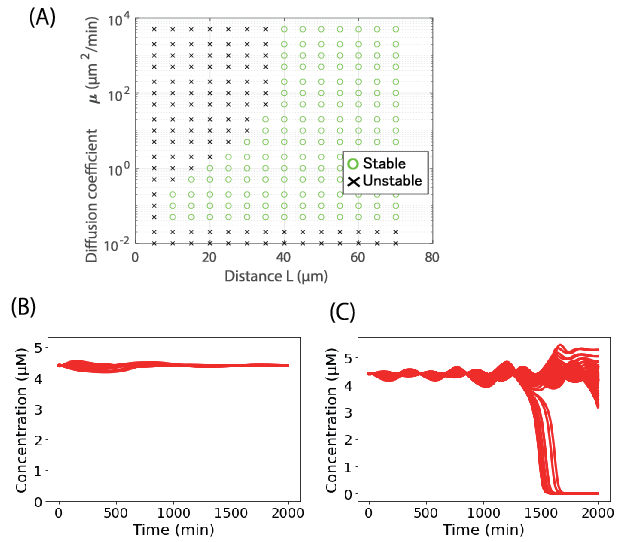}
    \caption{(A) The parameter map for the stability of the circulant MC system $\mathcal{M}$ for different communication lengths $L$ and diffusion coefficients $\mu$. (B) The behavior of the concentration of the repressor molecule in the reaction system $\Sigma_i$ around the spatially homogeneous equilibrium point for $i=1,2,\cdots,100$ when $\mu=10\,\si{\micro m^2\cdot min^{-1}}$ and $L=70\,\si{\micro m}$, and (C) when $L=10\,\si{\micro m}$. }
    \label{fig:map}
\end{figure}

We consider the circulant MC system illustrated in Fig. \ref{fig:ard}, where each nanorobot has an activator-repressor-diffuser (ARD) genetic circuit \cite{yutaka2015coordinated}. %
The function $\bm{f}(\bm{x}_i)$ of the reaction system is
\begin{equation}
    \bm{f}(\bm{x}_i) = \left[\begin{matrix}\displaystyle
        -\delta_{a}x_{i,1} + \gamma_{a}\frac{x_{i,1}^2}{K_{a}^2 + x_{i,1}^2}\frac{K_{r}^2}{K_{r}^2 + x_{i,2}^2}\\\displaystyle
        -\delta_{r}x_{i,2} + \gamma_{r}\frac{x_{i,1}^2}{K_{a}^2 + x_{i,1}^2}\frac{K_{d}^2}{K_{d}^2 + x_{i,3}^2}\\\displaystyle
        -\delta_{d}x_{i,3} + \gamma_{d}\frac{K_{r}^2}{K_{r}^2 + x_{i,2}^2} + k(x_{i,4} - x_{i,3})\\
        k_p(x_{i,3} - x_{i,4})
    \end{matrix} \right],\label{eq:f}
\end{equation}
where $x_{i,1}(t)$, $x_{i,2}(t)$, $x_{i,3}(t)$, and $x_{i,4}(t)$ are the concentrations of activator, repressor, signal molecule in the $i$-th nanorobot, and signal molecule outside of the nanorobot, respectively. The parameters $\delta_l$, $\gamma_l$, and $K_l$ are the degradation rate, the production rate, and the Michaelis Menten constant of the $l$-th molecular species, respectively. The parameter $k_p$ is the membrane transport rate. The average length of the communication channel is $L$, and the diffusion coefficient is $\mu$. The parameter values are shown in Table \ref{tab:param}, whose order of magnitude is consistent with widely used values for numerical simulations in synthetic biology \cite{yutaka2015coordinated, Basu2005,Li2021}. %
In what follows, we analyze the stability of the circulant MC system around the spatially homogeneous equilibrium point $\bm{x^{*}}=[5.09, 4.39, 8.08, 8.08]\,\si{\micro M}$ for different values of the communication length $L$ and the diffusion coefficient $\mu$.

\begin{table}
\begin{center}
\caption{Parameter sets for the analysis of the model in Fig. \ref{fig:mcsystems}, where each nanorobot has an ARD genetic circuit.}
\label{tab:param}
\begin{tabular}{ccc}\hline\hline
Parameter&&Value\\ \hline\hline
Production rate of $x_{i,1}(t)$&$\gamma_1$&$2.5\,\si{\micro M\cdot min^{-1}}$\\ 
Production rate of $x_{i,2}(t)$&$\gamma_2$&$2.68\,\si{\micro M\cdot min^{-1}}$\\ 
Production rate of $x_{i,3}(t)$&$\gamma_3$&$0.9\,\si{\micro M\cdot min^{-1}}$\\ 
Degradation rate of $x_{i,1}(t)$&$\delta_1$&$0.07\,\mathrm{min^{-1}}$\\ 
Degradation rate of $x_{i,2}(t)$&$\delta_2$&$0.07\,\mathrm{min^{-1}}$\\
Degradation rate of $x_{i,3}(t)$&$\delta_3$&$0.009\,\mathrm{min^{-1}}$\\ 
Dissociation constant of $x_{i,1}(t)$&$K_1$&$11\,\si{\micro M}$\\ 
Dissociation constant of $x_{i,2}(t)$&$K_2$&$9\,\si{\micro M}$\\ 
Dissociation constant of $x_{i,3}(t)$&$K_3$&$11\,\si{\micro M}$\\ 
Membrane transport rate&$k_p$&$0.5\,\si{min^{-1}}$\\ \hline\hline
\end{tabular}
\end{center}
\end{table}

\par
\smallskip
We consider $n=100$ nanorobots. The transfer function matrix of the MC channel $G(s)$ is obtained as the $100\times 100$ circulant matrix, and the reaction system is $H(s)=h(s)I$, where $h(s)=\bm{C}(sI-A)^{-1}\bm{B}$. The Jacobian matrix $A$ around the spatially homogeneous equilibrium point $\bm{x^{*}}=[5.09, 4.39, 8.08, 8.08]\,\si{\micro M}$ is obtained as
\begin{equation}
    A = \left[\begin{matrix}
    0.0453  & -0.0312  & 0 & 0\\
    0.0993  & -0.0700  & -0.0266 &  0\\
    0 & -0.0636 & -0.5900 & 0.5000\\
    0 & 0 & 0.5000 & -0.5000
    \end{matrix}\right],
\end{equation}
which leads to the transfer function %
\begin{equation}
    h(s) = \frac{10s^3 + 6.1469s^2 + 0.128s + 0.0003}{s^4 + 1.1147s^3 + 0.0701s^2 + 0.0003s}.
\end{equation}
The roots of the equation $\mathrm{den}(h(s))=0$ are computed as $-1.0480$, $0.0018+0.0217j$, $0.0018-0.0217j$, and $-0.0704$. Since the equation $\mathrm{den}(h(s))=0$ has two roots whose real part is positive, the reaction system $h(s)$ is unstable.
Using Theorem 1, the stability of the circulant MC system can be analyzed by computing the roots of the characteristic equation (\ref{eq:chat}) for $i=1,2,\cdots,51$.

\par
\smallskip
We then use Lemma 2 to analyze whether all the roots of the characteristic equation (\ref{eq:chat}) lie in the OLHP of the complex plane. 
Since the reaction system $h(j\omega)$ has two roots with positive real part, namely $P=2$, if the Nyquist plot of $-\lambda_i(j\omega)h(j\omega)$ makes two anti-clockwise encirclements of the point -1 and does not pass through the point -1 in the complex plane as the frequency $\omega$ increases for $i=1,2,\cdots,51$, the circulant MC system $\mathcal{M}$ is locally asymptotically stable, otherwise unstable.

\smallskip
\par
Fig. \ref{fig:map} (A) shows the parameter map for the stability of the circulant MC system for $n=100$ for different communication lengths $L$ and diffusion coefficients $\mu$. Fig. \ref{fig:map} (B) shows the concentration behavior of the repressor $x_{i,2}(t)$ in 100 nanorobots for $L=70\,\si{\micro m}$ and $\mu=10\,\si{\micro m^2\cdot min^{-1}}$ when a small perturbation $\bm{\eta}_r\in\mathbb{R}^{100}$ generated by random values is added to $\bm{\bar{x}}_4(0)\coloneqq[x_{1,4}(0),x_{2,4}(0),\cdots,x_{100,4}(0)]$, whose $i$-th entry is $x_{i,4}(0)=8.08\,\si{\micro M}$. The other molecular concentrations $x_{i,1}(0)$, $x_{i,2}(0)$, and $x_{i,3}(0)$ are not perturbed around $\bm{x}^*$ for all $i$. We can see that all molecular concentrations converge to the spatially homogeneous equilibrium point $\bm{x}^*$ corresponding to the analytical result in Fig. \ref{fig:map}. On the other hand, Fig. \ref{fig:map} (C) shows the concentration behavior of the repressor $x_{i,2}(t)$ in 100 nanorobots for $L=10\,\si{\micro m}$ and $\mu=10\,\si{\micro m^2\cdot min^{-1}}$ when a small perturbation $\bm{\eta}_r$ inputs to $\bm{\bar{x}}_4(0)$. We can see that the molecular concentrations transition to a different equilibrium point than $\bm{x}^*$. Thus, Fig. \ref{fig:map} verifies the proposed stability analysis method, which is helpful for the design of stable multi-agent MC systems, where the nanorobot population can achieve a cooperative behavior by synchronizing their state.

\par
\smallskip
As seen in Fig. \ref{fig:map}, the circulant MC system becomes stable as the communication length $L$ is greater. This result implies that the variation of the molecular concentration from the spatially homogeneous equilibrium point induced by a perturbation flows out into the large MC channels, thereby facilitating the convergence of the variation of the molecular concentration to zero. 
Also, as the diffusion coefficient $\mu$ increases, the MC system becomes unstable. 
This is because, when the diffusion coefficient $\mu$ is large, the variation of the molecular concentration can reach the neighboring nanorobots, which amplifies the variation of the molecular concentration in the nanorobot. 
On the other hand, for extremely small $\mu$, the variation of the molecular concentration in the nanorobot cannot be released into the MC channel, and thus the variation increases gradually, which leads to the unstable MC system. 
These properties are consistent with those of the quorum sensing mechanism \cite{waters2005quorum}, such as the property that a greater $L$ stabilizes the MC system, meaning that the system with a lower density of nanorobots can easily transition to another state.

\subsection{Stability analysis based on spatial frequency}
\label{sec:frequency}

\begin{figure}
    \centering
    \includegraphics[width=0.99\linewidth]{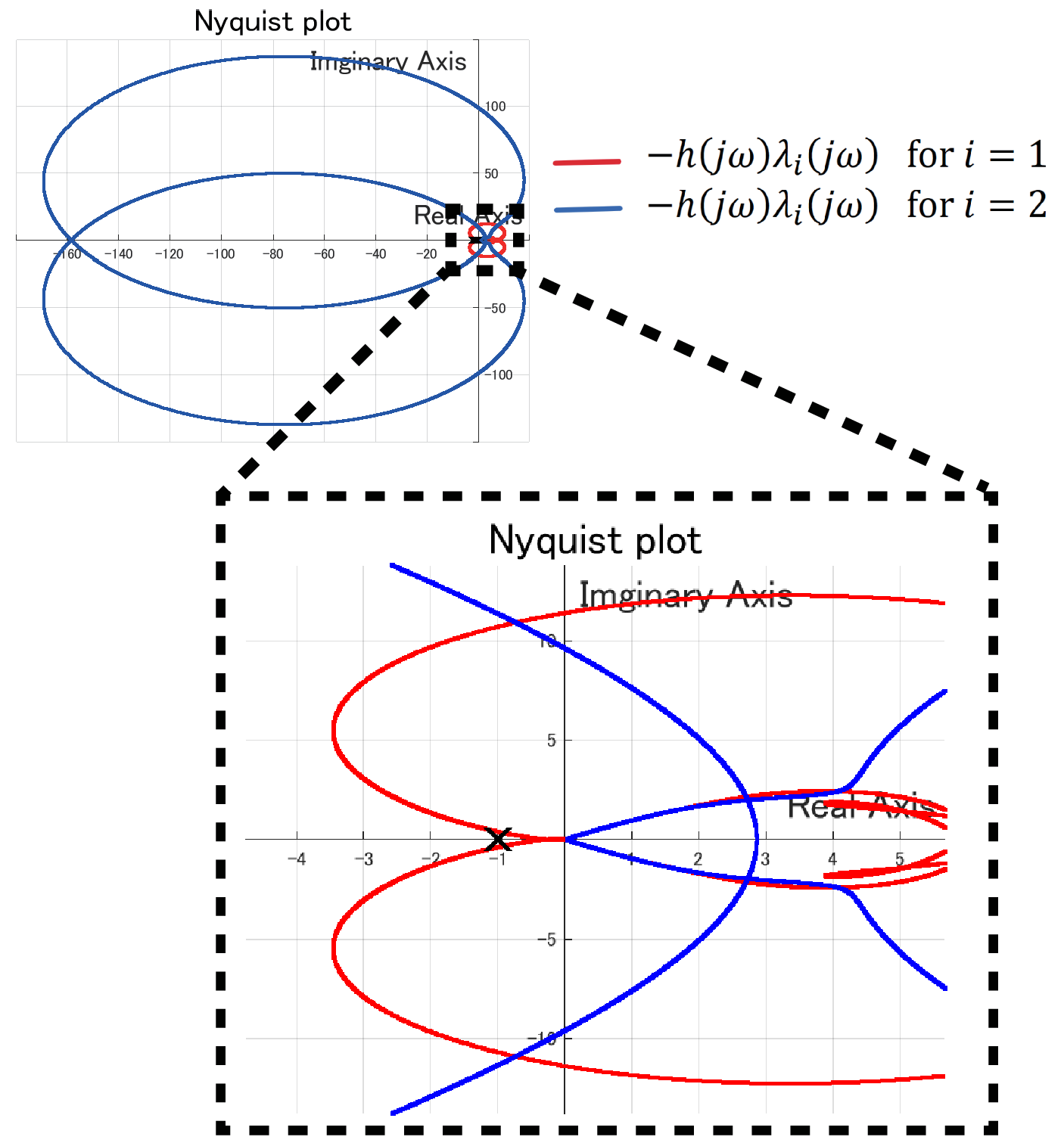}
    \caption{The Nyquist plot of the open-loop transfer function $-\lambda_i(j\omega)h(j\omega)$ of the subsystems $\hat{\mathcal{M}}_i$ for $i=1,2$ with the communication length $L=10\,\si{\micro m}$ and the diffusion coefficient $\mu=10\,\si{\micro m^2\cdot min^{-1}}$. The cross mark represents the point -1. }
    \label{fig:subnyquist}
\end{figure}

We next analyze the stability of the circulant MC system based on the spatial frequency to show that the convergence of the state of the circulant MC system to the spatially homogeneous equilibrium point depends on the frequency component of the small perturbation added to the system at $t=0$.

\par
\smallskip
For demonstration purpose, we consider the circulant MC system with $n=3$ nanorobots illustrated in Fig. \ref{fig:ard} with the parameter values in Table \ref{tab:param}. Figure \ref{fig:subnyquist} depicts the Nyquist plot of the open-loop transfer function $-\lambda_i(j\omega)h(j\omega)$ of the subsystem $\hat{\mathcal{M}}_i$ for $i=1,2$ when the diffusion coefficient $\mu=10\,\si{\micro m^2\cdot min^{-1}}$ and the communication length $L=10\,\si{\micro m}$. The figure shows that the Nyquist plot of $-h(j\omega)\lambda_1(j\omega)$ (red line) does not encircle the point -1 while the Nyquist plot of $-h(j\omega)\lambda_2(j\omega)$ (blue line) encircles the point -1 twice. Thus the subsystem $\hat{\mathcal{M}}_1$ is unstable while the subsystem $\hat{\mathcal{M}}_2$ is stable, which leads to the circulant MC system $\mathcal{M}$ unstable. This result implies that, although the system $\mathcal{M}$ is unstable, the state of the circulant MC system converges to the spatially homogeneous equilibrium state when the perturbation has no Fourier components of spatial frequency $\xi_1$.  

\begin{figure}
    \centering
    \includegraphics[width=0.9\linewidth]{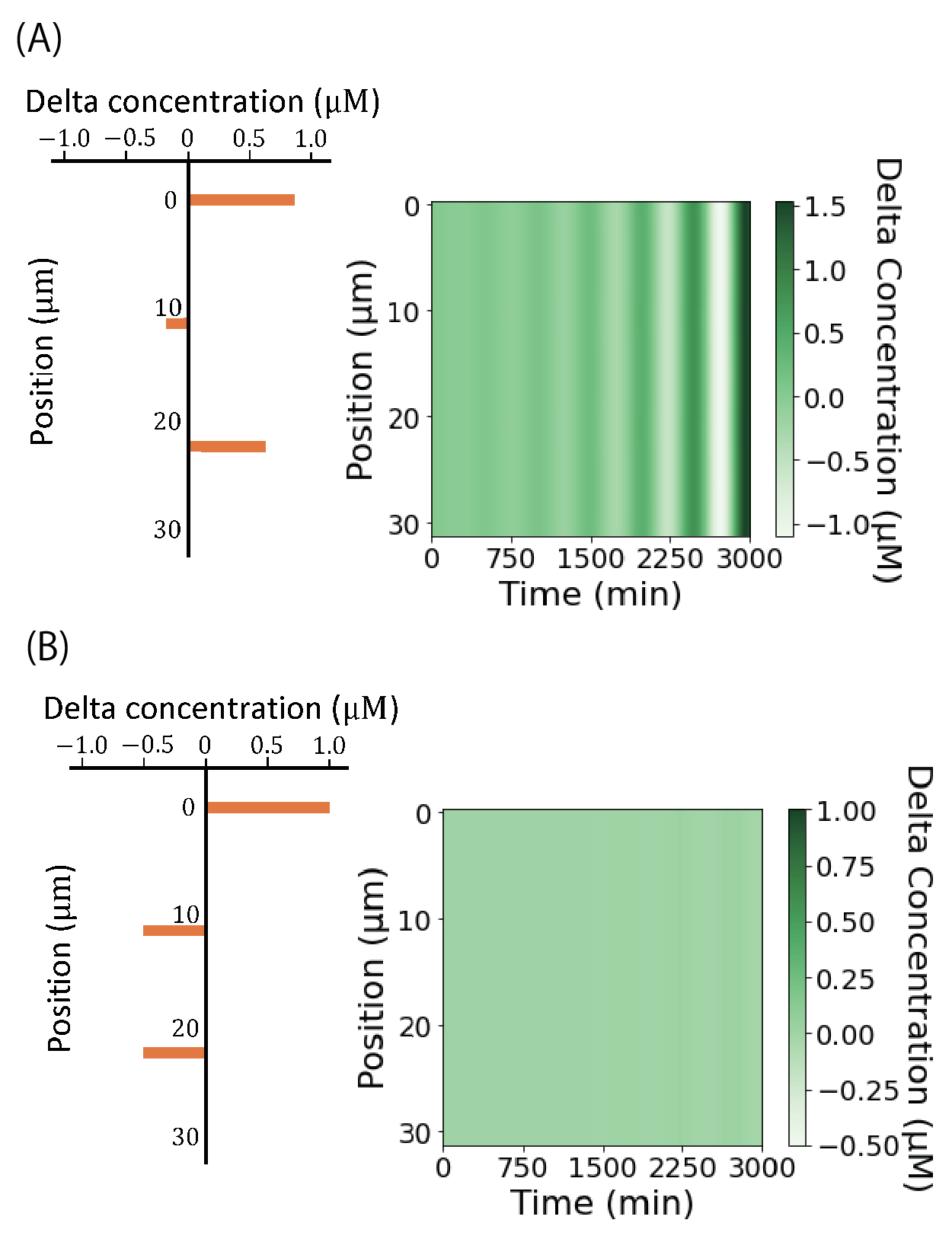}
    \caption{(A) (Left) The spatial distribution of the perturbation $\bm{\eta}_r$, which inputs to $\bm{\Tilde{y}}(0)$, and (Right) the time series data of the distribution of the concentration variation of the signal molecule $\Tilde{c}_i(t,r)$ for all $i$. (B) (Left) The spatial distribution of the perturbation $\bm{\eta}_c$, which inputs to $\bm{\Tilde{y}}(0)$, and (Right) the time series data of the distribution of the concentration variation of the signal molecule $\Tilde{c}_i(t,r)$ for all $i$.}
    \label{fig:timefreq}
\end{figure}

\par
\smallskip
Fig. \ref{fig:timefreq} shows the time series data of the variation of the concentration of the signal molecule for different initial perturbations around the spatially homogeneous equilibrium point $x^*_4$, which is $\Tilde{c}_i(t,r)=c_i(t,r)-x_4^*$ for all $i=1,2,3$
In particular, the concentration $\bm{\Tilde{x}}_4(0)\coloneqq[\Tilde{x}_{1,4}(0),\Tilde{x}_{2,4}(0),\Tilde{x}_{3,4}(0)]$ is perturbed by $\bm{\eta}_r\in\mathbb{R}^3$ whose $i$-th entry is $\eta_{ri} = \cos{(\xi_2iL)}$ in Fig. \ref{fig:timefreq} (A) and by $\bm{\eta}_c = [1.0, -0.5, -0.5]$ in Fig. \ref{fig:timefreq} (B), where the $i$-th entry of $\bm{\Tilde{x}}_4(0)$ is $\Tilde{x}_{i,4}(0)\coloneqq x_{i,4}-x_4^*$. The other molecular concentrations $x_{i,1}(0)$, $x_{i,2}(0)$, and $x_{i,3}(0)$ are not perturbed around $\bm{x}^*$ for all $i$. When the perturbation $\bm{\eta}_r$ inputs, the concentration variation $\Tilde{c}_i(t,r)$ diverges %
as seen in Fig. \ref{fig:timefreq} (A), implying that the molecular concentration transitions to another equilibrium point. On the other hand, when the perturbation $\bm{\eta}_c$ is added to the system, the concentration variation $\Tilde{c}_i(t,r)$ converges to 0 for all $i$ as seen in Fig. \ref{fig:timefreq} (B). These results arise because the perturbation $\bm{\eta}_r$ includes the Fourier component with the spatial frequency $\xi_1$ while the perturbation $\bm{\eta}_c$ does not include the Fourier component with $\xi_1$, which corresponds to the analytical result. Figures \ref{fig:subnyquist} and \ref{fig:timefreq} confirm that the molecular concentration either converges to the spatially homogeneous equilibrium point or transitions to another state depending on the spatial frequency of the perturbation. This phenomenon is very similar to the mechanism of Turing pattern formation, suggesting that the proposed method can be used to analyze spatial frequency-based biological phenomena such as Turing pattern formation.

\section{Conclusion}
\label{sec:conclusion}

In this paper, we have modeled large-scale multi-agent MC systems as circulant MC systems using transfer functions based on a diffusion equation. We have then proposed the method to analyze the stability for the circulant MC systems by decomposing the transfer function matrix of the circulant MC system into multiple single-input and single-output (SISO) systems. The proposed method can analyze the response of multi-agent MC systems for the spatial pattern of perturbation, which potentially leads to the analysis of spatial frequency-based biological phenomena such as Turing pattern formation. 
Finally, we have demonstrated the proposed method by analyzing the stability of a specific large-scale multi-agent MC system.

\par
\smallskip
In practice, the communication length $L$ might not be the same for all MC channels. Hence, our future work will be devoted to analyzing the effect of the nonuniform length $L$ on the stability of the MC system, in which case the communication matrix $G(s)$ is no longer circulant. 
Toward this extension, robust stability analysis for the multi-agent systems \cite{Kim2011} would be a useful approach. 

\appendix
\section{Proof of Lemma 2}
\label{sec:lemma2proof}

For the proof of Lemma 2, we first introduce the Nyquist stability criterion and then show the equivalence between the Nyquist stability criterion and Lemma2. Consider the characteristic equation (\ref{eq:chat}), and let $\hat{P}$ denote the number of the roots with positive real part of the equation $\mathrm{den}(h(s))\mathrm{den}(\lambda_i(s))=0$, where %
\begin{equation}
    \mathrm{den}(\lambda_i(s)) = \sqrt{\frac{\mu}{s}}\sinh{\left(\frac{L}{\sqrt{\mu}}\sqrt{s}\right)}\label{eq:denlambda}.
\end{equation}
The Nyquist stability criterion \cite{skogestad2005multivariable} states that all roots of the characteristic equation (\ref{eq:chat}) lie in the OLHP of the complex plane if and only if the Nyquist plot of the open-loop transfer function $-\lambda_i(j\omega)h(j\omega)$ makes $\hat{P}$ anti-clockwise encirclements of the point -1 and does not pass through the point -1 in the complex plane. 

\par
\smallskip
Next, we show $\hat{P}=P$. Let $\mathbb{Y}$ denote the solution set of the equation $\mathrm{den}(h(s))\mathrm{den}(\lambda_i(s))=0$. The solution set $\mathbb{Y}$ is equivalent to the set of the solution satisfying the equation $\mathrm{den}(h(s))=0$ or the equation $\mathrm{den}(\lambda_i(s))=0$. Since $\mathrm{den}(\lambda_i(s))=0$ has no roots, the solution set $\mathbb{Y}$ is equivalent to the solution set of the equation $\mathrm{den}(h(s))=0$. Therefore, the number of the roots with positive real part of the equation $\mathrm{den}(h(s))\mathrm{den}(\lambda_i(s))=0$ equals the number of the roots with positive real part of the equation $\mathrm{den}(h(s))=0$, {\it i.e.,} $\hat{P}=P$.


\end{document}